\documentclass[journal]{IEEEtran}
\ifCLASSINFOpdf
\else
\fi

\usepackage{url}
\usepackage{xcolor}
\usepackage{amsmath}
\usepackage{amsfonts}
\usepackage{multicol}
\usepackage{booktabs}
\usepackage{fancyref}
\usepackage{filecontents}

\usepackage{pgfplots}
\usepackage{tikz}
\usetikzlibrary{matrix,calc,shapes,spy}
\usepgfplotslibrary{fillbetween}

\usepackage{readarray}

\definecolor{mycolor1}{RGB}{0,0,0} % black
\definecolor{mycolor2}{RGB}{228,26,28} % Red
\definecolor{mycolor3}{RGB}{55,126,184} % blue
\definecolor{mycolor4}{RGB}{77,175,74} % green
\definecolor{mycolor5}{RGB}{152,78,163} % purple
\definecolor{mycolor6}{RGB}{255,127,0} % orange

\ifCLASSOPTIONcompsoc
  \usepackage[caption=false,font=normalsize,labelfont=sf,textfont=sf]{subfig}
\else
  \usepackage[caption=false,font=footnotesize]{subfig}
\fi

\begin{document}

\title{Stochastic Hybrid Approximation for\\Uncertainty Management in Gas-Electric Systems}

\author{Conor~O'~Malley,~\IEEEmembership{Member,~IEEE,}
        Gabriela~Hug,~\IEEEmembership{Member,~IEEE}
        and~Line~Roald,~\IEEEmembership{Member,~IEEE}% <-this % stops a space
%\thanks{thanks1}% <-this % stops a space
%\thanks{thanks2}% <-this % stops a space
\vspace{-1cm}
}

\markboth{Journal of \LaTeX\ Class Files,~Vol.~14, No.~8, August~2015}%
{Shell \MakeLowercase{\textit{et al.}}: Bare Demo of IEEEtran.cls for IEEE Journals}


\maketitle

\begin{abstract}

Gas-fired generators, with their ability to quickly ramp up and down their electricity production, play an important role in managing renewable energy variability. However, these changes in electricity production translate into variability in the consumption of natural gas, and propagate uncertainty from the electric grid to the natural gas system. 
To ensure that both systems are operating safely, there is an increasing need for coordination and uncertainty management among the electricity and gas networks.  
A challenging aspect of this coordination is the consideration of natural gas dynamics, which play an important role at the time scale of interest, but give rise to a set of non-linear and non-convex equations that are hard to optimize over even in the deterministic case. Many conventional methods for stochastic optimization cannot be used because they either incorporate a large number of scenarios directly or require the underlying problem to be convex. 
To address these challenges, we propose using a Stochastic Hybrid Approximation algorithm to more efficiently solve these problems and investigate several different variants of this algorithm. In a case study, we demonstrate that the proposed technique is able to quickly obtain high quality solutions and outperforms existing benchmarks such as Generalized Benders Decomposition. We demonstrate that coordinated uncertainty management that accounts for the gas system can significantly reduce both electric and gas system load shed in stressed conditions. 
\end{abstract}

% \begin{IEEEkeywords}
% gas-electric network, gas fired power plant, uncertainty, stochastic hybrid approximation.
% \end{IEEEkeywords}

% \fi
\IEEEpeerreviewmaketitle

\section{Introduction}
Natural gas fired power plants (GFPP) have become an increasingly important generation source due to low carbon emissions, low cost and high flexibility \cite{IEA}, and play an important role in managing variability and uncertainty from renewable energy resources. However, adapting GFPP output to balance variable RES leads to variations in the natural gas withdrawals from the gas pipeline system, thus propagating the uncertainty and variability associated with RES generation to the gas network.
Failure to consider this interdependence, particularly if coupled with generally high gas demand and significant level of uncertainty, can lead to sub-optimal performance or undesirable outages.  
Some electrical system operators include simplified representations of the gas network in their operational models \cite{FERC_ER20}.
However, accounting for the interdependence through simplified or sequential models may overly restrict operation \cite{omalley2020natural}. 
Furthermore, not accounting for the impact of uncertainty in both electric and natural gas systems can cause safety violations or disruption in gas supply to generators \cite{roald2020uncertainty}.
To address these issues, 
this paper develops a method for scheduling interconnected gas electric systems under uncertainty.

Coordination between gas and electric systems is particularly challenging because they operate on very different time scales. 
While electric systems almost immediately settle into a new steady state, 
the gas system typically operates in a transient dynamic state and operational decisions need to account for the slowly evolving system state
through the use of predictive models \cite{omalley2018security}. On the other hand, natural gas pipelines have a lot of inherent storage, frequently referred to as linepack, which allows the system to operate in a temporarily unbalanced state and helps buffer against uncertainty from the GFPP gas demand \cite{clegg2015integrated}. To assess the impact of RES uncertainty and gas-electric interdependency on intra-day operations, it is therefore necessary to model the natural gas dynamics.
The equations that govern the gas flow in pipelines are a set of partial differential equations \cite{osiadacz1984simulation} which can be discretized to give a non-convex optimization problem. This problem can be solved with local solvers \cite{zlotnik2016coordinated, mak2016efficient}, but remains computationally challenging. A common way of simplifying the problems is to use a steady state approximation based on the Weymouth equations, which is also non-linear and non convex, but can be approximated using piecewise linear approximations \cite{correa2014gas} or sequential linear programming \cite{lohr2020optimal,chaudry2008multi}. A number of convex relaxations have also been proposed for the steady state problem, including linear cutting planes \cite{tomasgard2007optimization}, second order cone constraints \cite{schwele2019coordination, wang2017convex} or semi-definite programming \cite{manshadi2018coordinated}. The resulting solutions provide a bound on the objective function, but not necessarily a solution that is feasible for the original model and thus, does not imply that the transient dynamic gas model will be feasible. 
Therefore, this paper consider the full transient dynamic model.

However, considering the gas system dynamics introduces some challenges, particularly in combination with consideration of uncertainty. The transient dynamic equations are both non-convex, meaning that many approaches for stochastic programming cannot be applied, and computationally intensive to include in an optimization problem, which effectively limits the number of scenarios that can be considered simultaneously. 
Techniques for stochastic problems can be applied to the convexified gas problem such as sample average approximation (SAA) methods \cite{qadrdan2013operating} which has good in-sample performance but can have a poor out of sample performance \cite{bertsimas2018robust}. Other methods include bound tightening via back off constraints \cite{liu2020dynamic}, adding a safety margin \cite{chen2019stochastic} or chance constrained optimization \cite{ratha2020affine}. Distributionally robust optimization has also be applied \cite{wang2019convex} typically with affine approximations and iterative strategies \cite{zhang2020distributionally,zhang2019two}. These methods do not require solving the problem for a large scenario set but require linear approximations of a non-linear system. This linearization is applied to all realizations in an uncertainty set, which may be inaccurate. 
The literature on optimization of dynamic natural gas systems under uncertainty is very limited. 
In \cite{zavala2014stochastic}, an SAA problem featuring the non-convex natural gas problem (without the electric system) is successfully solved for a small system and sample set, however the scalability of this problem is identified as an issue. 
In \cite{roald2020uncertainty}, a monotonicity property of dynamic gas networks from \cite{zlotnik2016monotonicity} was leveraged to formulate a robust gas-electric optimization problem based on two worst-case scenarios. An important challenge to this method is conservativeness, which arises from the limited conditions under which the monotonicity property holds. 

In this paper, we overcome these drawbacks by applying a stochastic approximation method. Related algorithms have previously been used for online optimization of reactive power in distribution grids \cite{kekatos2014stochastic}, in two stage stochastic problems with affine control policies \cite{kannan2020stochastic} and for power generation dispatch \cite{de2016adaptive}.
Specifically, we propose to use a stochastic hybrid approximation (SHA) \cite{cheung2000shape} which starts with a simple approximation that is iteratively updated using stochastic subgradients from a limited sample set.  
This allows us to solve the gas-electric problem for one scenario at a time, thus greatly improving computational tractability.
In \cite{de2016adaptive}, it is shown that the certainty equivalent problem, which replaces the uncertain variables with their expected values, provides a good initial approximation for power system dispatch. This choice of initial approximation is equivalent to the proximal stochastic gradient method \cite{nemirovski2009robust}, but with a non-euclidean norm.
The main contribution of this paper are as follows. First, we formulate the gas-electric problem under uncertainty as a two-stage stochastic optimization problem, where the second stage corresponds to operation under different wind generation scenarios. Second, we discuss the application of stochastic approximation algorithms to this problem. This includes a standard convex approximation, an approximation based on the certainty equivalent and a new algorithm based on what we refer to as the extrema equivalent (which incorporates the scenarios with highest and lowest wind generation). The advantages compared to previous approaches is the tractability for large scenario sets and the use of the true model for the gas network dynamics without any relaxations. The requirements for practical implementation such as suitable stopping criteria and parameter choices are also investigated.
Third, we demonstrate the performance of the SHA algorithm relative to benchmark algorithms on the non-convex gas and electricity problem, and demonstrate the importance of accounting for the gas system in dispatching decisions. 

The remainder of this paper is divided as follows. In section II, the two-stage optimization model of the gas-electricity network is presented.
In section III, we discuss the different solution algorithms, while Section IV presents numerical results. Section V summarizes and concludes.

\section{Modeling}
We next present our formulation for electric grid scheduling with consideration of (i) wind generation uncertainty (though other types of renewable generation or load, as well as uncertainty in the gas loads could easily be incorporated) and (ii) availability of natural gas for gas-fired power plants. While renewable energy uncertainty is treated as a random variable, the availability of natural gas is determined by whether or not the natural gas system is operating in a secure state. 
The electric grid operator thus aims to determine a generation dispatch which provides sufficient flexibility to respond to wind power uncertainty, while keeping the gas network safe and minimizing the expected cost of operation. 
We model this problem as a two-stage stochastic program, where the first stage problem chooses a generation dispatch schedule and the second stage problem redispatches generation to meet a particular wind power realization. 
The two stage optimization problem can be expressed as 
\begin{alignat}{2}
\quad \quad & \underset{x\in \mathcal{X}} 
{\text{Min.}}\quad &&  f(x)+\underset{\omega \in \Omega}{\mathbb{E}} \Big[ g(x,\omega)\Big] 
\label{mod:2stage_1st}
\end{alignat}
where $x$ is the first stage variables, i.e. the nominal generation dispatch, that belongs to a feasible set $\mathcal{X}$ and $f(x)$ is the costs associated with the first stage. The second part of the objective represents the expected cost of second stage operation ${\mathbb{E}}_{\omega \in \Omega} \left[ g(x,\omega)\right]$, where the operator decides on adjustments to generation, gas supply and load to satisfy system constraints given a realization of the wind power generation $\omega$ that comes from a set of i.i.d. scenarios $\Omega$, each of which describes the output for each wind farm across the time horizon of the problem. 
The second stage cost $g(x,\omega)$ depend on the first stage variables $x$ and the realization of the wind power generation $\omega$, and can be expressed as
\begin{alignat}{2}
g(x,\omega)=\quad \quad & \underset{y\in \mathcal{Y}(x,\omega)} 
{\text{Min.}}\quad &&  h(y).
\label{mod:stage_2nd}
\end{alignat}
Here, $y$ is the second stage variables representing real-time operation of the electric and gas systems, and the set $\mathcal{Y}(x,\omega)$ represents the system constraints which are a function of $x$ and $\omega$. The cost function $h(y)$ is the cost associated with the second stage. Next we provide details of the first and second stage models, including a detailed description of the electrical and gas network modeling.

\subsection{First stage problem}
In the first stage problem, the system operator determines a nominal schedule for the generators, denoted by $p_{g,t}$ for all generators $g \in \mathcal{G}$ and timesteps $t \in \mathcal{T}$. The generation output is bounded by the minimum $\underline{p}_{g,t}$ and maximum $\overline{p}_{g,t}$ operating limits and a ramping limit $r_g$, 
\begin{alignat}{2}
     &\underline{p}_{g,t} \leq p_{g,t} \leq \overline{p}_{g,t}, \quad &&\forall g \in \mathcal{G},t. \label{eq:DCOPF_Pmaxmin}\\
   &| p_{g,t-1} - p_{g,t}| \leq r_g, \quad &&\forall g \in \mathcal{G},t \in \{t_2,\ldots, T\}, \label{eq:DCOPF_ramp}
\end{alignat}
The first stage objective minimizes the cost of the initial generator dispatch,
\begin{alignat}{2}
& f(x)=\sum_{t \in \mathcal{T}} \sum_{g \in \mathcal{G}} C_g p_{g,t}
\end{alignat}
where $C_g$ is the marginal cost of generation. In summary, the first stage variables comprise $x=\{p_{g,t}\}$ and are constrained by $\mathcal{X}=\{\eqref{eq:DCOPF_Pmaxmin}- \eqref{eq:DCOPF_ramp}\}$.
Note that the first stage problem does not enforce load balance or network constraints because the available wind is still unknown. The system constraints, including both the physical system models (power and gas flow) as well as technical limits (line flows and pressure limits), are enforced in the second stage.

\subsection{Second stage problem}
The second stage optimizes the system operation, given the first stage decision $x$ and a wind power realization $\omega$. 
All variables associated with specific wind power realization are denoted by the subscript $\omega$. The second stage includes models for the electric and gas networks along with coupling constraints that link the two networks (i.e., through the gas consumption of GFPPs). We first describe the electric network, then the gas system and then the coupling constraints.

\subsection{Electric Network Modelling}
In the following, we present the multi-period DC optimal power flow (OPF) problem that models the electric system. This model is standard in the literature \cite{lohr2020optimal,chaudry2008multi,schwele2019coordination}, though we extend it by introducing a load adding variable to ensure relative complete recourse in the presence of ramping constraints. 

\subsubsection{Generation constraints} In the second stage, we obtain a realization of the wind generation scenario and redispatch the generators to balance the net load. Denoting the scheduled generation by $p_{g,t,\omega}$, the first and second stages are linked by
\begin{subequations}
\begin{align}
     p_{g, t, \omega} =p_{g, t} \;   \quad \forall, g \in \mathcal{G},t,\omega \label{eq:fix_lambda}
\end{align}
The generation redispatch is represented by the non-negative variables $p^{+}_{g,t,\omega}\geq 0$ and $p^{-}_{g,t,\omega}\geq 0$. The total generation after redispatch must respect the generation and ramping limits,
\begin{align}
 & \underline{p}_g\leq  p_{g, t, \omega} + p^{+}_{g,t,\omega} -p^{-}_{g,t,\omega} \le  \overline{p}_g, \;\;\; &&\forall g \in \mathcal{G},t,\omega, \label{eq:DCOPF_prout} \\
& 0 \leq p^{+}_{g,t,\omega} \leq r_g -(p_{g,t,\omega} -p_{g,t-1,\omega})
, \;\;\; &&\forall g \in \mathcal{G},t,\omega, \label{eq:DCOPF_prmax} \\
&  0 \leq p^{-}_{g,t,\omega} \leq r_g +(p_{g,t,\omega} -p_{g,t-1,\omega})
. \;\;\; &&\forall g \in \mathcal{G},t,\omega, \label{eq:DCOPF_prmin} 
\end{align}
The last term in \eqref{eq:DCOPF_prmax} and \eqref{eq:DCOPF_prmin} represent the reduction in ramping capability due to ramping in the first stage. 

\subsubsection{Power flow constraints} 
The power flow for each line in the network $\ell \in \mathcal{L}$ is modelled with the DC approximation and is denoted $f_{\ell,t,\omega}$.
The flow is expressed as 
\begin{align}
     f_{\ell, t, \omega} = B_{\ell} \sum_b A(\ell,b) \theta_{b, t, \omega}, 
     \quad     \forall \ell \in \mathcal{L},t,\omega,\label{eq:DCOPF_flowdef} 
\end{align}
where $B_{\ell}$ is the line susceptance, $A(\ell,b)$ is the incidence matrix of the network and $\theta_{b,t,\omega}$ represents the phase angles at each node. The power flow is limited according to the thermal limits $\overline{f}_{\ell}$ of the line, and a reference angle $\theta_{b_1, t, \omega}$ is fixed to avoid degeneracy. 
\begin{alignat}{2}
     -\overline{f}_{\ell} \leq f_{\ell, t, \omega} \leq \overline{f}_{\ell}, &
     \quad     \forall \ell \in \mathcal{L},t,\omega,\label{eq:DCOPF_flowdef} \\
      \theta_{b_1, t, \omega}=0,  &\quad \forall t,\omega, \label{eq:DCOPF_ref}
\end{alignat}

\subsubsection{Nodal power balance}
The net load comprises a fixed load $L_{b,t}$ at each bus $b \in \mathcal{B}$ and the output $W_{j,t,\omega}$ of the wind farms $j \in \mathcal{W}$, which takes on a different value in each scenario. We also include variables which allow the system operator to spill wind power $w^{\text{spill}}_{j,t,\omega}$ or shed load $l^{\text{shed}}_{b,t, \omega}$ at a high cost. 
The load shedding and wind spill are constrained by the total available load and total available wind power, 
\begin{alignat}{2}
0 \leq& l^{\text{shed}}_{b,t, \omega} \le  L_{b,t},  
\quad \forall b\in \mathcal{B},t,\omega, \label{eq:DCOPF_shedlim} \\
0 \leq& w^{\text{spill}}_{j,t, \omega} \le  W_{j,t, \omega}, 
\quad \forall j \in \mathcal{W},t,\omega,. \label{eq:DCOPF_spilllim} 
\end{alignat}

In addition to the variables for wind spill and load shed, we include an auxillary variable $l^{\text{add}}_{b,t, \omega}$ which represents load adding (i.e. an increase in the load at a node). 
This variable (along with $w^{\text{spill}}_{j,t,\omega}$ and $l^{\text{shed}}_{b,t, \omega}$) allows us to ensure relative complete recourse for the electric system, i.e., that the second stage model admits a feasible solution regardless of the first stage decision. This characteristic is a necessary feature for our iterative solution algorithm presented below, and is not satisfied without $l^{\text{add}}_{b,t, \omega}$ due to the presence of ramping constraints and non-zero minimum generation limits. We note that $l^{\text{add}}_{b,t, \omega}$ is only non-zero at intermediate iterations, and always is expected to become zero in the final solution. 
We also enforce that the load adding is non-negative and does not exceed a sufficiently large upper bound $L^{\text{add}}_{b,t}$ which guarantees that any excess generation can be absorbed, 
\begin{alignat}{2}
0 \leq& l^{\text{add}}_{b,t, \omega} \leq L^{\text{add}}_{b,t},  
\quad \forall \in \mathcal{B},t,\omega, \label{eq:DCOPF_addlim} 
\end{alignat}

Given these definitions, the power balance for each node is expressed as 
\begin{alignat}{1}
&\!\!\!\sum_{g \in \mathcal{M}^{\mathcal{G}}_b} \!\!\left( p_{g,t,\omega}\!+\!p^{+}_{g,t,\omega}\!-\!p^{-}_{g,t,\omega}\right) \!+\!\!\! 
 \sum_{j \in \mathcal{M}^{\mathcal{W}}_b} \!\!\left(W_{j,t,\omega}\!-\!w^{\text{spill}}_{j,t,\omega} \right)\!=\!\nonumber\\
& ~~L_{b,t} \!-\! l_{b,t,\omega}^{\text{shed}} 
\!+\!l_{b,t,\omega}^{\text{add}}\!+\! \sum_{\ell} A(\ell, b) f_{\ell,t,\omega},
 \quad \forall b \in \mathcal{B},t,\omega, \label{eq:DCOPF_bal} 
\end{alignat}
\end{subequations}
where $\mathcal{M}^{(\cdot)}_b$ maps the elements of $(\cdot)$ to bus $b$.

\subsubsection{Electric network objective}
The second stage objective of the electricity network minimizes cost of generation redispatch. We assume that redispatch decisions are more costly than the nominal dispatch, i.e. $C_g^{-}\leq C_g \leq C^{+}_g$. We also assume a high penalty for both load shedding $C^{\text{VoLL}}$ and load adding $C^{\text{add}}$, giving the following objective function
\begin{alignat}{2}
 h^{\text{elec}}(y^{\text{elec}})=&\sum_{t} \Big[\sum_{g}\Big( C_{g}^{+} p^{+}_{g,t,\omega}\nonumber
 -  C_{g}^{-} p^{-}_{g,t,\omega} \Big) \\
 &+ \sum_b ( C^{\text{VoLL}} l_{b,t, \omega}^{\text{shed} }+ C^{\text{add}} l_{b,t, \omega}^{\text{add} }) \Big]\label{eq:SS_DCOPF_obj} 
\end{alignat}
where $y$ represents the second stage electrical variables for scenario $\omega$. These variables include $y_{\omega}^{\text{elec}}=\{$ 
$p_{g,t,\omega}$,
$p^{+}_{g,t,\omega}$,
$p^{-}_{g,t,\omega}$,
$w^{\text{spill}}_{j,t,\omega}$,
$l^{\text{shed}}_{b,t,\omega}$,
$l^{\text{add}}_{b,t,\omega}$,
$f_{\ell,t,\omega}$, 
$\theta_{b,t,\omega}\}$ and are constrained by $\mathcal{Y}_{\omega}^{\text{elec}} =\{\eqref{eq:fix_lambda} -\eqref{eq:DCOPF_spilllim} \}$.

\subsection{Gas Network Modeling} 
The second stage gas flow problem is similar to the models in \cite{zlotnik2016coordinated, roald2020uncertainty}, and identifies gas supply and compressor setpoints to satisfy the demand and network constraints while minimizing the expected cost. The demand includes consumption of normal loads and GFPPs, which in turn depends on the wind scenario $\omega$. 
We assume that there are no ramping constraints on the compressors or supply points limiting their ability to adjust their dispatch points. Because the system is able to fully adjust to the wind power realization, 
the natural gas system only appears in the second stage. We further assume that the gas system problem satisfies relatively complete recourse if we include load shedding of the non-GFPP load in the network (i.e., the system is able to sustain all possible GFPP demands if all other gas load is shed).

\subsubsection{Gas Network Dynamics}
The isothermal transient flow of gas in a pipeline network is governed by the partial differential equations (PDEs) that describes the relationship between the varying pressure $\pi$ and mass flow rate $m$. The constants $V_m,V_p,V_f$ are determined based on the pipe geometry.
\begin{subequations}
\label{eq:pde}
\begin{align}
    \frac{\partial \pi}{\partial t} + V_{m} \frac{\partial m}{\partial t}&=0\\
    \frac{\partial m}{\partial t} + V_{p} \frac{\partial \pi}{\partial x}&=-V_f \frac{m|m|}{\pi}
\end{align}
\end{subequations}

To model the gas flow equations, we adopt a discretize-then-optimize approach whereby the PDEs are approximated in space and time using finite difference methods and the resulting equations are included as constraints in the optimization problem. 
First the gas network is described as a directed graph comprising an original set of nodes $\mathcal{N}_0$ and pipes $\mathcal{P}_0$. A spatial discretization of the pipelines is performed by introducing auxiliary nodes, denoted by $\mathcal{N}_{a}$, between the original nodes $\mathcal{N}_{0}$ that subdivides the original pipe into subpipes, denoted by $\mathcal{P}$. The discretized gas network therefore comprises the set of the subpipes $\mathcal{P}$, and set of all nodes $\mathcal{N}=\mathcal{N}_0 \cup \mathcal{N}_{a}$. 
The time discretization of the gas problem is chosen to match the time discretization in the OPF problem. 

The state variables at each time step $t$ and scenario $\omega$ of the gas network are the pressure $\pi_{i,t,\omega}$, and the mass flows in and out of each subpipe $m_{i,t,\omega}$ which are defined at each node $i \in \mathcal{N}$. 
Each subpipe has an associated mass flow $m_{ij,t,\omega}$ and a pressure $\pi_{ij,t,\omega}$, expressed as the average of the values at each end of the pipe 
 \begin{subequations}
\begin{alignat}{2}
m_{ij,t,\omega}&=0.5(m_{i,t,\omega}+m_{j,t,\omega}) \quad \forall (i,j) \in \mathcal{P}, t,\omega, \label{eq:OGF_mavg}\\
\pi_{ij,t,\omega}&=0.5(\pi_{i,t,\omega}+\pi_{j,t,\omega}) \quad \forall (i,j) \in \mathcal{P}, t,\omega. \label{eq:OGF_pavg}
 \end{alignat}
Applying the finite difference methods to the PDEs for each pipe subsection $\forall (i,j) \in \mathcal{P}, t  \in T,\omega$ results in 
\begin{align}
& u_t\textstyle{\frac{\left(\pi_{ij,t,\omega}-\pi_{ij,t-1,\omega}\right)}{\Delta t}}\!+\!V_m\textstyle{\frac{m_{j,t,\omega}-m_{i,t,\omega}}{\Delta x}} \!=\!0 \\%\notag\\
 &u_t\textstyle{\frac{\left(m_{ij,t,\omega}-m_{ij,t-1,\omega}\right)}{\Delta t}} \!+\!\textstyle{V_f\frac{m_{ij,t,\omega}|m_{ij,t,\omega}|}{\pi_{ij,t,\omega}}} %\notag\\
 \!+\!\textstyle{V_p \frac{\pi_{j,t,\omega}-\pi_{i,t,\omega}}{\Delta x}}\!=\!0  \label{eq:OGF_momentum} 
\end{align}
where $\Delta x$ is the spatial discretization, i.e. the length of the subpipe, and $\Delta t$ is the temporal discretization. The parameter $u_t=1$ for all time, except for $t_1$ where $u_{t_1}=0$ to remove the temporal derivative. This implies that the gas network starts from steady state operation at $t_1$. 

\subsubsection{Pressure and compressor constraints}
These pressure constraints can be expressed as
\begin{align}
\underline{\pi}_{i} \leq \pi_{i,t,\omega} \leq& \overline{\pi}_{i}, \quad \forall i \in \mathcal{N},t,\omega  \label{eq:OGF_plim}  \\
\pi_{i_1,t,\omega} =& \hat{\pi}_{i_1}, \quad \forall t,\omega  \label{eq:OGF_p_ref}  
\end{align}
where the lower and upper limits on the pressure $\underline{\pi}_{i},\overline{\pi}_{i}$ represents contractual and safety limits on the gas pressure and \eqref{eq:OGF_p_ref} defines a reference pressure $\hat{\pi}_{i_1}$ to remove degeneracy.
The gas network also contains a set of compressors $\mathcal{C}$ which increase the output pressure by a multiplicative factor of $\kappa_{ij,t,\omega}$, giving rise to the following constraints
\begin{align}
    \pi_{j,t,\omega}=\kappa_{ij,t,\omega}\pi_{i,t,\omega} \quad \forall (i,j) \in \mathcal{C},t,\omega \label{eq:OGF_comp}\\
 \underline{\kappa}_{ij} \leq \kappa_{ij,t,\omega} \leq \overline{\kappa}_{ij}, \quad  \forall (i,j) \in \mathcal{C},t,\omega    \label{eq:OGF_clim}
 \end{align}
where $\underline{\kappa}_{ij},\overline{\kappa}_{ij}$ are compressor upper and lower bounds.
Compressors consume either gas or electricity in order to provide the compression. 
However, the amount of gas or electricity consumed is negligible relative to the overall system consumption, and is omitted in the network model. To incentivize low energy compression, we instead penalize the compressor use directly in the objective function. 

\subsubsection{Gas supply constraints} The natural gas supply is denoted by $s_{s,t,\omega}$ for each of the supply points $s \in \mathcal{S}$, and is limited according to 
\begin{align}
    \underline{s}_{s}   \leq s_{s,t,\omega} \leq \overline{s}_{s}, \quad \forall s \in \mathcal{S},t,\omega    
\end{align}
where $\underline{s}_{s}$ and $\overline{s}_{s}$ represent the minimum and maximum supply. 

\subsubsection{Gas balance constraints}
The load in the natural gas network comprises the nodal non-GFPP demand $D_{n,t,\omega}$ and the nodal GFPP demand $d_{g,t,\omega}$ for each GFPP $g \in \mathcal{G}_g \subseteq \mathcal{G}$. Similar to the electric network, we allow for shedding of gas load $d^{\text{shed}}_{n,t,\omega}$. 
Given these definitions, the mass balance for each node $\forall n \in \mathcal{N}, t,\omega$ is expressed as 
\begin{align}
&\sum_{s \in \mathcal{M}_n^{\mathcal{S}}} \!\!\!s_{s,t,\omega}\!=\!
D_{n,t,\omega} \!-\! 
d^{\text{shed}}_{n,t,\omega}\!+\!
\!\!\!\sum_{g \in \mathcal{M}_n^{\mathcal{G}_g}} \!\!\!d_{g,t,\omega}\nonumber\\
&\quad +\!\!\!\sum_{i \in \mathcal{M}_n^{\text{in}}} \!\!\!m_{i,t,\omega}\!
-\!\!\!\sum_{i \in \mathcal{M}_n^{\text{out}}}\!\!\!m_{i,t,\omega}\! 
+\!\!\!\sum_{i \in \mathcal{M}_n^{\text{in,c}}} \!\!\!m^{\text{c}}_{i,t,\omega}\!-\!\!\!\sum_{i \in \mathcal{M}_n^{\text{out,c}}}\!\!m^{\text{c}}_{i,t,\omega}\label{eq:OGF_bal}
\end{align}
Here, $\mathcal{M}_n^{\text{in}},~\mathcal{M}_n^{\text{out}}$ represent the set of pipes with mass flows in and out of the node $n$, $m^{\text{c}}_{i,t,\omega}$ is the mass flow through compressor $i$ and  $\mathcal{M}_n^{\text{in,c}},~\mathcal{M}_n^{\text{out,c}}$ represent the set of flows into and out of node $n$ through the compressors.

The natural gas within the pipelines, typically referred to as the linepack, act as an inherent storage buffer and allows the network to operate in an unbalanced state for a period of time. 
The optimization problem could reduce the cost of supplying gas by fully depleting the linepack, which is a bad starting point for operation beyond the considered optimization horizon. To avoid this unwanted depletion of the linepack, we enforce that 
the total supply over the duration of the optimization horizon is balancing the total demand in that same time period,
\begin{align}
\sum_{t} \Big[\sum_{s} s_{s,t,\omega}
\sum_{n } (D_{n,t,\omega}-
d^{\text{shed}}_{n,t,\omega})
\sum_{g \in \mathcal{G}_g } d_{g,t,\omega}\Big]=
0, \forall \omega  \label{eq:OGF_linepack}    
\end{align}
 \end{subequations}
 
\subsubsection{Gas network objective}
The gas network objective minimizes the cost of the gas supplied, while penalizing any load shedding and compressor usage. This is expressed as
\label{mod:OGF}
\begin{alignat}{2}
h^{\text{gas}}(y^{\text{gas}})\!=\! &\sum_{t}\!
  \Big[\sum_{s} C_s  s_{s,t,\omega}\! + \!\!\!\! \sum_{(i,j) \in \mathcal{C}} \!\!\!C_{ij} \kappa_{ij,t,\omega}\! +\!\sum_n C^{\text{sh}} d^{\text{shed}}_{n,t,\omega}  \Big] \label{eq:OGF_obj}
\end{alignat}
where $C_{s}$ is the marginal cost of supply, $C^{\text{sh}}$ is the cost of gas load shedding and $C_{ij}$ is the cost of compression. The second stage gas variables for scenario $\omega$ comprises $y_{\omega}^{\text{gas}}=\{$
$s_{s,t,\omega}$,
$\kappa_{ij,t,\omega}$, 
$m_{ij,t,\omega}$,
$m_{i,t,\omega}$,
$\pi_{ij,t,\omega}$,
$\pi_{i,t,\omega}$, 
$d_{g,t,\omega}$,
$d^{\text{shed}}_{n,t,\omega}$,
$m^{\text{c}}_{i,t,\omega} \}$ 
and are constrained by $\mathcal{Y}_{\omega}^{\text{gas}} =\{\eqref{eq:OGF_mavg} -\eqref{eq:OGF_linepack} \}$

\subsection{Network Coupling} 
The electrical and gas network are coupled by the constraints that link the GFPP power in the electrical network to its gas demand in the gas network. This relationship is described by the heat rate curve,
\begin{align}
    d_{g,t,\omega} = \eta_g \left(p_{g,t,\omega}+p^{+}_{g,t,\omega}-p^{-}_{g,t,\omega} \right)  \quad \forall g \in \mathcal{G}_g,t,\omega
    \label{eq:coupling_eq}
\end{align}
where the heat rate $\eta_g$ determines how much gas is needed to produce one unit of electricity. 
The electric system cost \eqref{eq:SS_DCOPF_obj} implicitly includes the cost of natural gas to the GFPPs, which is also included in the gas network cost \eqref{eq:OGF_obj}. 
To avoid double counting in the combined problem, we subtract the cost of gas from the costs of operating the gas network. We assume that the GFPPs have a fixed cost for gas $C_g$, that could arise from a bilateral contract, and there are no additional O\&M costs associated with GFPP operation. The full second stage objective is then given by 
\begin{alignat}{2}
h(y_{\omega})=  &  h^{\text{elec}}(y_{\omega}^{\text{elec}})+ h^{\text{gas}}(y_{\omega}^{\text{gas}})- \sum_{t}   \sum_{g \in \mathcal{G}_g} C_{g} \frac{ d_{g,t,\omega}}{ \eta_g}
\end{alignat}
and the full second stage problem for scenario $\omega$ is
\begin{alignat}{2}
g(x,\omega)=  & \underset{y_{\omega}^{\text{elec}},y_{\omega}^{\text{gas}}}
{\text{Min.}} && h(y_{\omega}) \label{mod:Full_SS}
  %\nonumber
  \\
 & \text{s.t.} && \text{Eq.}\eqref{eq:coupling_eq}, %\\
 %&&&
 y_{\omega}^{\text{elec}} \in \mathcal{Y}_{\omega}^{\text{elec}}(x),
 y_{\omega}^{\text{gas}} \in \mathcal{Y}_{\omega}^{\text{gas}}(x) \nonumber
 \end{alignat}

\section{Solution Algorithms}
\label{sec:Sol_Alg}
The two-stage problem formulation described in the previous sections is a non-convex optimization problem which can be solved as single optimization problem for small scenario sets and networks. However, as the scenario set or network increases in size, the required computational effort or time may become impractical.
In this section we describe how to solve problem \eqref{mod:2stage_1st} using the Stochastic Hybrid Approximation method, a decomposition method which breaks the problem into subproblems that are easier to solve.

\subsection{Stochastic Hybrid Approximation}

Stochastic Hybrid Approximation (SHA) is an iterative algorithm that approximates the second stage cost, or the so-called the recourse function, $\mathbb{E}[ g(x,\omega)]$ by the combination of an initial function $ \hat{Q}_0(x)$ and an independent linear correction term with coefficient $\overline{\lambda}_i^{(\nu)}$ for each value of $x_i$. The coefficients $\overline{\lambda}^{(\nu)}$ are initially set to zero and is updated at each iteration $\nu$. The approximation at any iteration is:
\begin{align}
\label{eq:SHA_Approx}
    \underset{\omega \in \Omega}{\mathbb{E}} \Big[ g(x,\omega)\Big] \approx  \hat{Q}_0(x) 
    + \sum_i \overline{\lambda}_i^{(\nu)} x_i
\end{align}
leading to the overall problem approximation
\begin{alignat}{2}
S^{(\nu)}(x)=\quad \quad & \underset{x\in \mathcal{X}} 
{\text{Min.}}\quad &&  f(x)+ \hat{Q}_0(x) + \sum_i \overline{\lambda}_i^{(\nu)} x_i
\label{mod:2stage_1st_SHA}
\end{alignat}
In each iteration, we solve the approximation problem to produce a solution for the first stage variable $x$.
\begin{alignat}{2}
x={\text{argmin}}\quad &&  S^{(\nu)}(x)
\end{alignat}
The second stage approximation is then updated by solving the second stage problem $g(x,\omega)$  for a set of scenarios $\omega$, each of which can be solved independently and in parallel. In this paper, we will only use a single scenario in each iteration. 

The linear correction term $\overline{\lambda}_i$ for each first stage variable is updated using the stochastic subgradient $\lambda_i$. The value of $\lambda_i$ is the dual of \eqref{eq:fix_lambda}, which represents the sensitivity of the second stage objective value to the first stage variable. The overall update is expressed as  
\begin{align}
\label{eq:SHA_Update}
    \overline{\lambda}_i^{(v+1)}=\overline{\lambda}_i^{(\nu)}+\alpha^{(v)}\big(\lambda^{(v)}_i -( \hat{q}_{0i}^{(\nu)} + \overline{\lambda}_i^{(v)})\big) 
\end{align}
where $\hat{q}_{0i}^{(\nu)}$ is the derivative of the approximation   $\partial \hat{Q}_0(x)/\partial x_i$ evaluted at $x_i^{(\nu)}$ and the parameter $\alpha^{(v)}$ is a step size.

The step size impacts the speed of convergence, but there is no agreed procedure for choosing the step size apart from the guideline that it should decrease as the iterations increase, but not too rapidly \cite{yousefian2012stochastic,tan2016barzilai}. A common approach is a step size $\alpha^{{(v)}} = \rho/v$ where $\rho$ is a parameter that can be tuned to improve performance and we adopt this approach.

In addition to choosing a step size, we need to choose an initial approximation function $\hat{Q}_0(x)$. Simple approximations can lower the computation burden when solving the overall approximation \eqref{mod:2stage_1st_SHA}, but may require more iterations. Conversely, more accurate approximations can be improved with less iterations, but result in more challenging overall approximation problem. In the following, we present three different alternatives for the approximation function.

\subsubsection{Convex approximation}
In \cite{cheung2000shape}, a convex quadratic function is suggested as an approximation and can also be used in this problem. The approximation is expressed as
\begin{align}
\label{eq:SHA_Approx}
    \hat{Q}_0(x) = \sum_i a_{i} x_{i}^2+ b_{i} x_{i}
\end{align}
and the gradient $\hat{q}_{0i}^{(\nu)}$ is given by $\hat{q}_{0i}^{(\nu)} =2a_i x_i^{(\nu)}+b_i$.
This approximation requires an initial choice of $a_{i}$ and $b_{i}$. Choosing good values for $a_{i}$ and $b_{i}$ is non-trivial and can significantly impact convergence and the quality of the final result. 
Our approach is to choose a value for $a_i$ and calculate a corresponding value for $b_i$ such that the algorithm produces the dispatch that would occur if a single average wind scenario is used. We refer to this approximation as the stochastic hybrid approximation with convex approximation (SHACV).

\subsubsection{Adaptive Certainty Equivalent Approximation}
In the first iterations, the convex approximation often provides solutions that are very far away from optimal. The algorithm is improved upon in \cite{de2016adaptive} by observing that the certainty equivalent, i.e. dispatching the power systems assuming the expected values of the uncertain variables, can serve as a good initial approximation of the recourse function. The certainty equivalent approximation is the second stage problem for the average wind scenario and is expressed as
\begin{align}
   \hat{Q}_0(x) = \underset{\omega \in \Omega^{\text{CE}}}{\mathbb{E}}\big[g(x,\omega) \big] =g(x,\mathbb{E}[\omega])
   \label{eq:Q0_cvx}
\end{align}
with $\Omega^{\text{CE}}=\{\mathbb{E} [\omega] \}$. The problem solved at each iteration is 
\begin{alignat}{2}
\quad \quad & S^{(\nu)}(x)=\underset{x\in \mathcal{X}} 
{\text{Min.}}\quad &&  f(x)+\underset{\omega \in \Omega^{\text{CE}}}{\mathbb{E}}\big[h(y) \big]+ \sum_i \overline{\lambda}_i^{(\nu)} x_i\nonumber
  \\
 & \text{s.t.} && \text{Eq. }\eqref{eq:coupling_eq}  \label{mod:SHA_CVX_Full}   \quad \forall \omega \in \Omega^{\text{CE}}\\
 &&& y_{\omega}^{\text{elec}} \in \mathcal{Y}_{\omega}^{\text{elec}}(x) \quad \forall \omega \in \Omega^{\text{CE}}, \nonumber\\
 &&&\;y_{\omega}^{\text{gas}} \in \mathcal{Y}_{\omega}^{\text{gas}}(x)\quad \forall \omega \in \Omega^{\text{CE}}, \nonumber
\end{alignat}
We obtain $\hat{q}_{0i}^{(\nu)}$ in a similar way as $\lambda$, i.e. by solving $g(x,\mathbb{E}[\omega])$ and obtaining the dual of \eqref{eq:fix_lambda}.
By providing a better starting point for the algorithm, the certainty equivalent approximation can improve the convergence and reduce the need for tuning. However, we both need to solve a larger first stage problem as well as two second stage problems, the subproblem for a random wind scenario $\omega$ and the subproblem for $\mathbb{E}[\omega]$. 
We will refer to this approximation as the stochastic hybrid approximation with adaptive certainty equivalent (SHACE).

\subsubsection{Adaptive Extrema Equivalent Approximation}
A shortcoming of the certainty equivalent approximation is that the optimization problem typically will not initially schedule adequate flexibility to manage different wind realisations. 
We therefore propose to incorporate two scenarios that represent the extrema of the uncertainty in the first stage problem, namely the scenarios with the most and least total wind energy denoted by $\omega^{\text{max}}$ and $\omega^{\text{min}}$. By including these scenarios and their respective constraints, the need for flexibility is evident already in the first stage problem. The flexibility requirement is then tuned through the iterations of the algorithm. With this adjustment, the approximation becomes
\begin{align}
\label{eq:SHA_Approx}
    \hat{Q}_0(x) = \!\!\underset{\omega \in \Omega^{\text{EE}}}{\mathbb{E}}\big[g(x,\omega) \big] \!= \!\frac{1}{2}\!\left(g(x,\omega^{\text{max}})\!+ \!g(x,\omega^{\text{min}})\right)
\end{align}
where $\Omega^{\text{EE}}=\{\omega^{\text{max}},\omega^{\text{min}}\}$. The problem solved at each iteration is similar to \eqref{mod:SHA_CVX_Full}, but with  $\Omega^{\text{CE}}$ replaced by $\Omega^{\text{EE}}$.
We define $\hat{q}_{0i}^{(\nu)}=\frac{1}{2}\left(\lambda_i^{\text{max}}+\lambda_i^{\text{min}}\right)$ where $\lambda_i^{\text{max}}$ and $\lambda_i^{\text{min}}$ are the duals of constraint \eqref{eq:fix_lambda} in the subproblems of the respective wind scenarios $\omega^{\text{max}},\omega^{\text{min}}$.

Including the the two extreme scenarios can further reduce the required number of iterations, however it also increases the computational time per iteration.
This is because we must solve the subproblem three times, once for the randomly selected scenario $\omega$ and once for each of the extrema scenarios $\omega^{\text{max}},\omega^{\text{min}}$. 
This approximation will be referred to as the stochastic hybrid approximation with adaptive extrema equivalent (SHAXE).\\

\subsubsection{Stopping criteria and algorithm output}
The SHA algorithms produce a sequence of solutions $x^{(\nu)}$, which converges as $\nu$ goes to $\infty$ for convex problems. 
However, we solve a non-convex problem and use only a finite number of iterations. We therefore need to determine (i) a stopping criterion for the algorithm and (ii) what to return as the final solution. 

The true objective value for a given solution $x$ is obtained by solving the the second stage problem for all second stage scenarios. This is computationally costly and thus not practical to evaluate at each iteration.
The objective value of the SHA algorithm is not a good indicator, as
the SHA algorithm is trying to match the gradient of the approximation to the gradient of the original objective function rather than matching the objective value. Therefore, the objective value (and changes in the objective value) of the approximation \eqref{eq:SHA_Approx} may not correlate with the objective value of the original problem \eqref{mod:2stage_1st}.
Furthermore, because the algorithm is stochastic in nature, the true objective value is not monotonically decreasing as the number of iterations increase (i.e., $x^{(\nu+1)}$ is not necessarily a better solution than $x^{(\nu)}$).

We therefore suggest to use a weighted average of $x^{(\nu)}$ to determine an average solution $\overline{x}^{(\nu)}_n$, as proposed in \cite{nemirovski2009robust} and used in \cite{taheri2020strategic}.
The average is based on a sliding window of length $n$, where the window length determines the level of variability of the objective. We also use a weighting factor $1/\alpha^{(\nu)}$ to emphasize results from recent iterations. The weighted average value $\overline{x}^{(\nu)}_n$ at each iteration is defined as
\begin{align}
    \overline{x}^{(\nu)}_n =
    \textstyle{
    \frac{\sum_{i=i_0}^{\nu}x^{(i)}/\alpha^{(i)}}{\sum_{i=i_0}^{\nu}1/\alpha^{(i)}}
    }
\end{align}
where $i_0=\text{max}(1,\nu-n+1)$ to account for iterations shorter than the window length. 

We formulate a stopping criteria based on changes in this weighted average value $\overline{x}^{(\nu)}_n$, as smaller changes in $\overline{x}_n$ typically correlate with smaller changes in the original objective value. 
The stopping criteria is defined as when the average solution update $\Delta^{(\nu)}_n$ is less than a specified tolerance $\epsilon_{\text{tol}}$,  
\begin{align}
    \Delta^{(\nu)}_n =\textstyle{\frac{||\overline{x}^{(\nu)}_n-\overline{x}^{(\nu-1)}_n||}
    {||\overline{x}^{(\nu)}_n||}} \leq \epsilon_{\text{tol}}
    \label{eq:err_update}
\end{align}
Another option for terminating the algorithm is to set an iteration limit or a time limit. This approach can be beneficial in time critical applications but gives little insight on the progress of the algorithm.

\subsection{Benchmarking Algorithms}
We compare our proposed solution algorithm against two other methods for solving the problem.

\subsubsection{One Shot Solution} 
For small problems and sample sizes the problem can be solved as a single optimization problem which incorporates all the scenarios $\omega \in\Omega$ and expresses the expected second stage cost as $\mathbb{E}_{\omega \in\Omega}[g(x,\omega)]=\frac{1}{|\Omega|}\sum_{\omega \in\Omega} g(x,\omega).$ 

\subsubsection{Generalized Benders Decomposition}
The Generalized Benders Decomposition approximates the recourse function by successively adding cutting planes at the current solution of $x$, which leads to a piece-wise approximation  of $\mathbb{E}[ g(x,\omega)]$. Using a multi-cut approach, which adds multiple cutting planes at each iteration, the first stage problem becomes:
\begin{subequations}
\label{mod:Benders}
\begin{alignat}{2}
& \underset{x\in \mathcal{X}} 
{\text{Min.}}\quad && f(x) + \sum_{\omega}\eta_{\omega} \\ 
& \text{s.t.}&& \eta_{\omega} \geq g(x^{(v)},\omega) +  \sum_i \lambda_i^{(v)}(x_i-x_i^{(v)}) \quad \forall \omega,v
\end{alignat}
\end{subequations}
where $\eta_{\omega}$ is an auxiliary variable that represents the piece-wise approximation of the objective function from scenario $\omega$, while $g^{(v)}_{\omega}$ is the true second stage objective from iteration $v$ which was obtained with the first stage variables $x^{(v)}$ and scenario $\omega$. 

There are several drawbacks to the Generalized Benders Algorithm. First, it does not guarantee convergence for non-convex problems, as a non-convex problem can result in adding cuts that exclude part of the solution space \cite{sahinidis1991convergence}. 
Secondly, when the problem is non-convex there is no guarantee the upper and lower bounds calculated in the algorithm (which are typically used as a stopping criterion) are true upper and lower bounds. 
The termination criteria for the algorithm is therefore chosen to be when the changes in both of the bounds is less than a pre-specified tolerance $\varepsilon$. 

Third, this algorithm requires solving the second stage problem for every scenario $\omega$ at each iteration, as it is not known \textit{a priori} which of the scenarios will provide cuts that improve the approximation. There exists variations on the algorithm that can merge cuts to make the first stage problem more efficient \cite{vandenbussche2019data} or identify subproblems that provide improving cuts \cite{mazzibenders} if the subproblems are convex. However, here we only consider the basic algorithm. 

\subsection{Evaluating Performance}
We evaluate the performance of the algorithms by evaulating \emph{solution time} and \emph{total solution cost}. The total solution cost is evaluated by fixing the first stage variables and evaluating the first stage cost $f(x)$ and the expected value of the recourse function for a set of scenarios $\Omega^{\text{eval}}$. The solution quality is denoted as $V(x)$ and is expressed as 
\begin{align}
    V(x)=f(x)+\underset{\omega \in \Omega^{\text{eval}}}{\mathbb{E}} \Big[ g(x,\omega)\Big] 
\end{align}
The solution quality $V(x)$ can be evaluated both for the set of scenarios used in the optimization (which we will refer to as the training scenario set) and on previously unseen scenarios (referred to as the testing scenario set).

\section{Case Study}
In this section we investigate the performance of the proposed SHA algorithm for the coupled gas-electric problem. First, we introduce the test case and perform initial analysis on the algorithm performance and parameter choices. We then demonstrate the advantages of the proposed algorithm compared to Benders, and show why it is important to incorporate gas system constraints in the electric scheduling algorithm. The algorithms are implemented in the Julia language \cite{bezanson2017julia} using the JuMP package for optimization \cite{DunningHuchetteLubin2017}. The non-linear problems are solved using Ipopt \cite{wachter2006implementation}. However, to reliably obtain high quality dual values, each problem is subsequently linearized at the solution and solved again with Gurobi \cite{gurobi}.

\subsection{Test Case}
As a basis for our case study, we use the gas and electricity network that has previously been analysed in \cite{zlotnik2016coordinated,roald2020uncertainty} and is available at \cite{data24}. The cost of electrical load shedding is 1000 USD/MWh, gas load shedding is set to 5 USD/kg and the compression cost is set 1 USD. The generation quadratic cost functions is converted to a linear function with same total costs. The upwards and downwards redispatch costs are $C^{+}_g=1.05C_g$ and $C^{-}_g=0.94 C_g$ respectively. The total wind farm capacity is set to half of the nominal load and uniformly distributed among the possible locations. We use 100 wind scenarios from \cite{bukhsh2015integrated}, available at \cite{winddata}, to produce a wind generation profile that is used by all the windfarms (i.e., we assume full spatial correlation).
These scenarios are divided into a training set with 80 scenarios and a testing set with 20 scenarios to verify the result via an out-of-sample evaluation. 
If the SHA algorithms require more than 80 iterations, we reuse the same training scenarios but in a different (randomly assigned) order.

We solve the problem for the first 12 hours of the day, as this time horizon allows us to solve the problem as a one shot optimization problem, which is useful for benchmarking. This solution of the one shot problem will be denoted as $x^{\text{OS}}$.

\subsection{Algorithm parameter tuning}

First, we investigate the impact of the parameters on the performance of the SHA algorithm.

\subsubsection{Window Length}
\begin{figure}
\centering
\subfloat
    [Solution quality for different window lengths \label{fig:window_analysis}]
    {%
	\includegraphics[scale=0.8]{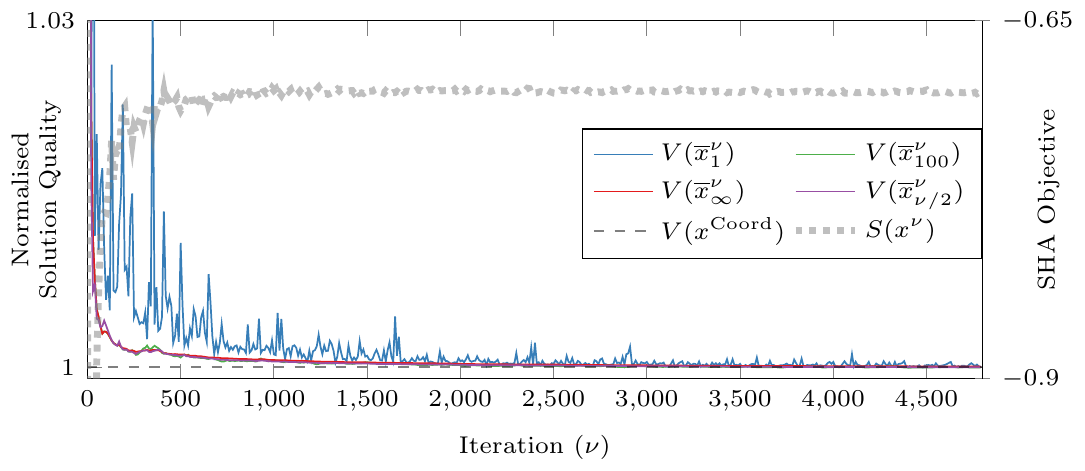}
    }
\\
\subfloat
    [Solution update for different window lengths\label{fig:err_analysis}]
    {
	\includegraphics[scale=0.8]{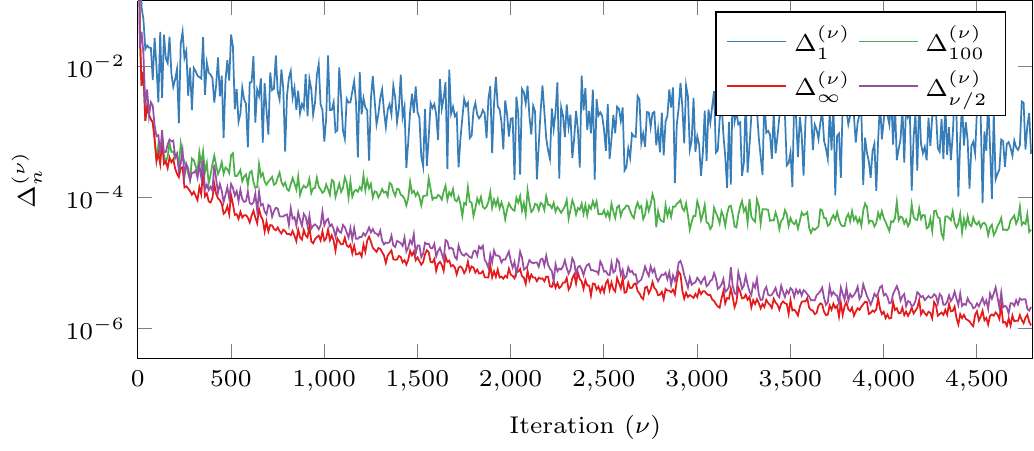}
    }
    \caption{Comparison of solution quality based on aposteriori evaluation (top) and the solution update stopping criterion (bottom) for SHACV (with $a=1000$, $\rho=1$) for different sliding window lengths $n=\{1, 100, \nu/2, \infty\}$. The top plot also shows the objective at each iteration (in grey).}
    \vspace{-0.5cm}
\end{figure}

The choice of a window length $n$ for the averaging of the solution determines how we obtain a final solution from the solution sequence produced by the SHA algorithm. We run the SHACV algorithm with $a=1000$ and $\rho=1$ for a fixed (large) number of iterations, and calculate the solution $\overline{x}^{\nu}_n$ using different window lengths $n=\{1,100,\infty,\nu/2\}$. These values correspond to no averaging ($n=1$), averaging across 100 iterations, and averaging across all or half of the iterations seen so far. The solution quality $V(\overline{x}^{\nu}_n)$ is calculated at each iteration $\nu$ using the training set of samples, and is normalized by the one-shot optimization solution $V(x^{\text{OS}})$. The results are shown in \Fref{fig:window_analysis}. 

We observe that all averages converge to the objective value of the one shot optimization problem $V(x^{\text{OS}})$, regardless of  the window length $n$. \Fref{fig:window_analysis} shows that a window length of $n=1$ (i.e. no averaging) exhibits the most variability in solution quality between iterations, making it difficult to choose a solution from the sequence produced from the SHA algorithm. 
Window lengths of $n>1$ reduce the variability and result in more stable improvement in solution quality. The differences in solution quality for $n=\{100,\infty,\nu/2 \}$ are minor in this case.

\subsubsection{Stopping Criteria}
\Fref{fig:window_analysis} also shows the objective value of the SHA problem $S(x^\nu)$ in each iteration. This value is independent of the averaging and is increasing with the number of iterations. This clearly demonstrates that changes in  $S(x^\nu)$ do not correlate well with the actual solution quality and is not a good stopping criteria. 
We therefore investigate the proposed stopping criterion $\Delta^{(\nu)}_n$, which is based on change in the solution update $\overline{x}^{\nu}_n$. For each window length $n$, the solution update $\Delta^{(\nu)}_n$ is calculated according to \eqref{eq:err_update} and shown in \Fref{fig:err_analysis}.

The solution update metric $\Delta^{(\nu)}_n$ shows a decreasing trend in all cases. When the window length is $n=1$, the decrease is relatively slow and there is more variation in the update. 
For larger window lengths there is a more pronounced decrease, and less variability. In all cases, the decreasing trend becomes less pronounced over time, but the solutions obtained with larger averaging reach smaller values.  

The decrease in this metric is expected due to the decreasing step size in the algorithm, and when the update becomes small, it can indicate that the solution is reaching convergence.
The metric can however also saturate when the sliding window length is not long enough to sufficiently smooth the solution and for this reason a value of $n=\infty$ gives the most consistent correlation between the update metric and the solution quality. We use $n=\infty$ for the remainder of this paper.

\subsubsection{Other parameters}
In addition to the choices of window length $n$ and stopping criteria, the other algorithmic parameters that must be chosen are values for stepsize $\rho$ and coefficient $a$ (for the convex approximation only). A smaller value for $\rho$ can better leverage a good initial approximation, but can slow convergence in the later iterations. A large value initially induces large variations in the solution quality until enough samples have been collected. We found that $\rho=1$ typically provides good results. 
The choice of $a$ is most dependent on the problem, and can influence the speed of convergence, but is only necessary for the convex approximation. 
For the sake of brevity, the parameter sweep analysis has been omitted.

\subsection{Benchmarking}
\begin{figure}
    \centering
	\includegraphics[scale=0.8]{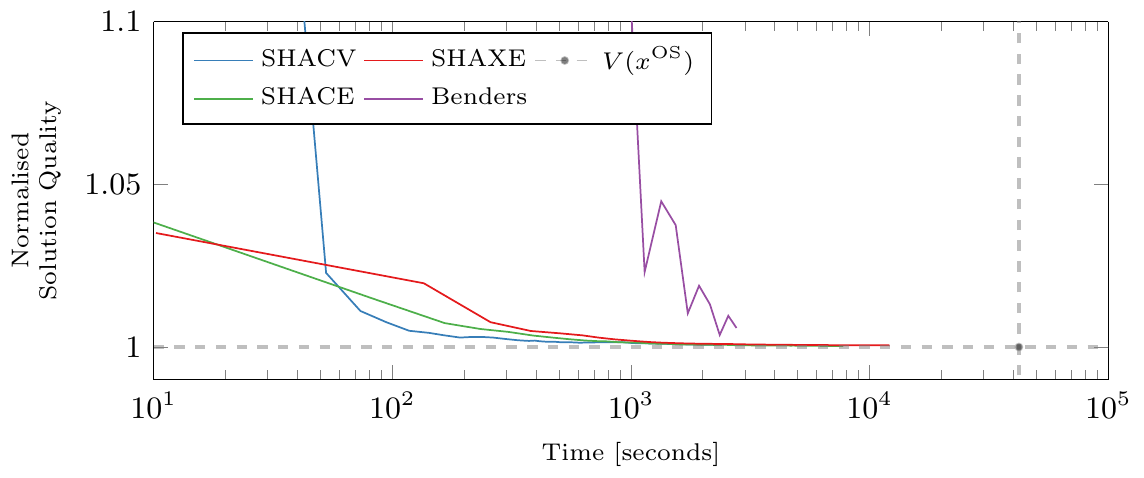}
	\includegraphics[scale=0.8]{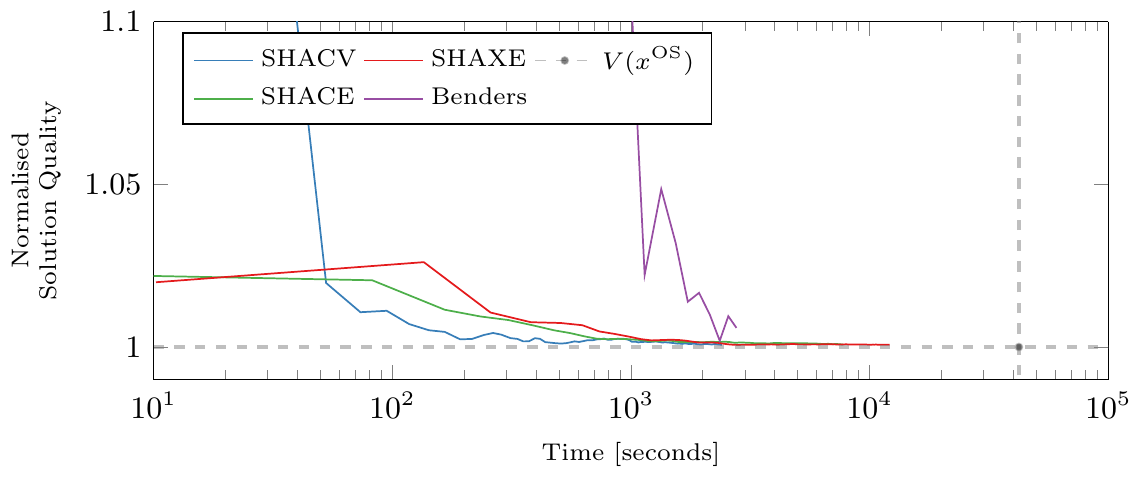}
     \vspace{-0.5cm}
    \caption{Comparison of solution quality for training (top) and testing (bottom) scenario sets for all algorithms on the OPGF problem. Ideally, the solution quality should be 1, and higher values indicate worse solutions.}
    \label{fig:opgf_compare_obj}
    \vspace{-0.5cm}
\end{figure}

Next we compare the different SHA algorithms to the Generalized Benders Decomposition and the one shot optimization, and compare solutions in terms of both solution quality and computational time. 
The SHA algorithms use a step size parameter $\rho=1$ and are run for a fixed number of iterations (4800 for SHA and 3200 for SHACE and SHAXE). The SHACV uses a coefficient $a=1000$. The Benders algorithm is run until there is a 1\% gap between the second stage approximation and the true second stage objective.
Figure \ref{fig:opgf_compare_obj} shows the solution quality of each algorithm against the time taken to achieve that solution quality.  In this plot, the solution quality for the training set is shown in the top plot while the solution quality for the previously unseen testing set is shown in the bottom plot. The solution quality is normalized against the objective values of the one shot optimization $V(x^{\text{OS}})$. A lower value (i.e., closer to 1) indicates a better solution.

In \fref{fig:opgf_compare_obj}, the one-shot optimization problem takes 11.7 hours to obtain a solution, as indicated by the horizontal dotted line. In comparison to this, the Benders algorithm took 0.5 hours to reach a 1\% gap. The SHACV, SHACE and SHAXE approximations reach a 1\% gap in 0.02 0.04 and 0.06 hours, and continue to improve after that. The results are consistent between both the training and testing scenario sets, indicating the robustness of this solution to unseen scenarios.

The SHA algorithms clearly need much less time to provide a solution of similar quality as compared with the benchmark algorithms. 
When comparing the SHA approximations to each other, it is not surprising that SHACV is faster than the other two. The SHACV only requires the solution of one the second stage problem, while the the certainty equivalent (SHACE) and extrema equivalent (SHAXE) approximations require the solution of two and three subproblems, respectively.

While all the SHA algorithms converge to the same value, there are some differences in the initial iterations. For SHACV, the solutions in the initial iterations are bad, then quickly improves. The other approximations contain a more detailed model of the problem and the initial iterations already produce solutions that are quite close to the optimal. 
Overall the SHACV has the best performance, but this was only achieved after extensive tuning of the parameters. Meanwhile, the SHACE and SHAXE algorithms are relatively insensitive to the tuning and still fast. We conclude that SHACE may provide the best trade-off between tuning and computational performance in many situations.

\subsection{Impact of gas network on power system operations.}

\begin{figure}
    \centering
\includegraphics[scale=0.8]{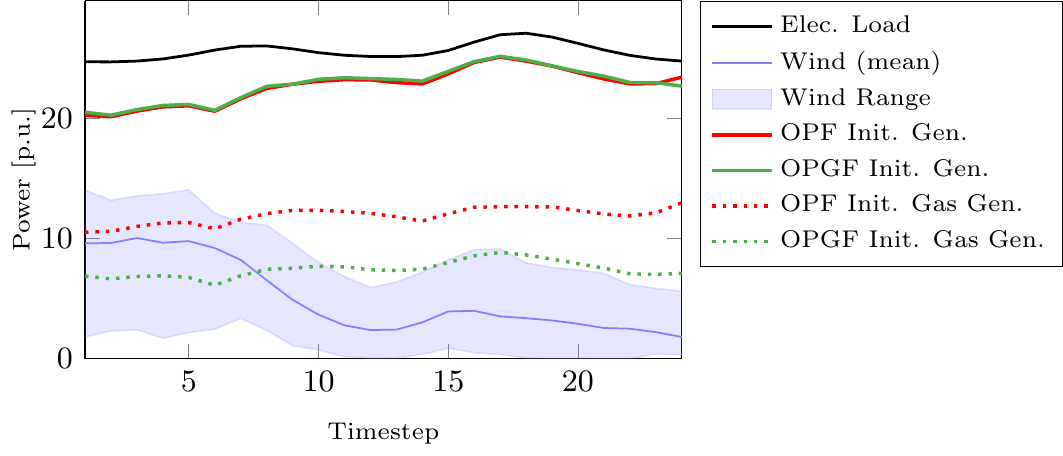}
    \vspace{-0.2cm}
        \caption{Evolution of the load, wind power and the initial (first stage) dispatch of generators and gas generators OPF and OPGF.}
    \label{fig:gen_compare}
   
\end{figure}

Finally, we assess the importance of coordinated gas-electric dispatch under uncertainty. We solve the two-stage optimal gas and power flow (OGPF) problem \eqref{mod:2stage_1st} using the SHACV algorithm. To emphasize the gas generation, the gas supply cost and GFPP costs are reduced to 25 \% of their original values while the non-GFFP costs are doubled. We compare the result with the solution of a two-stage OPF problem that represents the same electric system model, but does not include gas system constraints. This model is linear and can be solved directly using Gurobi. Both of these problems are solved using a 24 h time horizon and 80 training scenarios. The results are based on evaluating the same 80 training scenarios.

The temporal evolution of the electrical load and the range wind generation across all scenarios is shown in Fig. \ref{fig:gen_compare}. This figure also shows the first stage generation dispatch that results from the OPF and from the OPGF problem. The total generation in each case is similar, however the amount of gas generation (indicated by the dashed lines) is reduced when the gas constraints are considered. Notice that there is a difference between the total electric load and the first-stage generation dispatch. This difference is due to the fact that we do not require the total amount of generation to match the total load, because we do not know how much wind generation will be available. This difference is however made up for in the second stage, ensuring a balanced system.

\begin{figure}
    \centering
\includegraphics[scale=0.8]{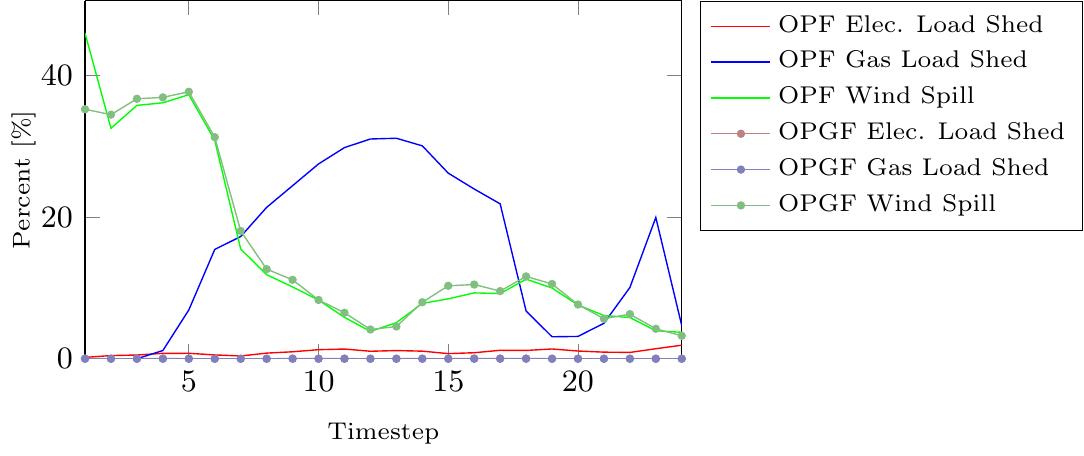}
    \vspace{-.8cm}
    \caption{Evolution of average electrical and gas load shedding and wind spill over the course of the day for the solutions obtained with the OPF and OPGF problems. All values are expressed as percentages of available wind and load.}
    \label{fig:mitigation}
\end{figure}

To compare the quality of first stage decisions obtained with and without considering the gas systems, we fix the first variables to the results from the OPF and OPGF problems and solve the second stage problem including both the electric and the gas network constraints. 
Generators can be redispatched in response to different realizations of the wind generation, but it may also be necessary to spill wind and shed load (both in the electrical and gas systems) to obtain a feasible solution. A summary of the wind spill and load shed values is given is given in Table \ref{tab:mitigation_breakdown} (top). 
The OPGF is able to avoid any gas load shed and only incurs a maximum electric load shed of 0.52\%. In contrast, the OPF sheds up to 5\% of the electric load, and more than 15\% of the gas load to maintain feasible operations. Figure \ref{fig:mitigation} shows the average mitigation actions at each time step (expressed as a percentage of total load and wind availability).  We observe that the wind spill is concentrated in the early hours of the day, when the load is low and the wind availability is high. 
The OPF solution causes electric and gas load shed in periods with low wind spill. The OPGF solution manages to keep feasible operations with almost no load shed (i.e., both gas and electric load shed are very close to zero throughout the day).

A breakdown of the first and second stage costs is given in Table \ref{tab:cost_breakdown} (bottom). We see that while the first stage costs are lower for the OPF than the OPGF, this is outweighed by a much higher second stage costs. When comparing the second stage cost for the OPF and the OPGF, we see that the electric system cost is higher for the OPGF, both on average and when considering the maximum and minimum values. However, the real difference is for the gas system cost, which is an order of magnitude larger for the OPF than the OPGF. 
This is because the first stage solution obtained with the OPF (i.e., without consideration of limitations in the gas system) leads to a large amount of non-GFPP load shedding in the second stage.

\begin{table}[]
\centering
\caption{Expected mitigation actions (top) and total system costs (bottom) given first stage OPF and OPGF results.}
\begin{tabular}{@{}l|ccc|ccc@{}}\toprule
\textbf{Average Value}        & \multicolumn{3}{c|}{ $x$ OPF} & \multicolumn{3}{c}{$x$ OPGF} \\ 
        & Mean    & Max.     & Min.    & Mean    & Max.   & Min.    \\ \midrule
Wind Spill  {[}\%{]}        &25.4&	41.86&	1.95&	25.39&	42.16&	2.19\\
Elec. Load Shed {[}\%{]}&0.95 &	4.99 &	0    &	0.01 &	0.52 &	0\\
Gas Load Shed {[}\%{]}       &14.95&	15.81&	14.46&	0    &	0    &	0\\
\bottomrule
\end{tabular}
\label{tab:mitigation_breakdown}
\vspace{5pt}
\begin{tabular}{@{}l|ccc|ccc@{}}\toprule
    \textbf{Cost} {[}Million CHF{]}      & \multicolumn{3}{c|}{ $x$ OPF} & \multicolumn{3}{c}{$x$ OPGF} \\ 
        & Mean    & Max.     & Min.    & Mean    & Max.   & Min.    \\ \midrule
1$^{\text{st}}$ Stage Elec.  & 0.95    & -       & -      & 1.09    & -  & -   \\
2$^{\text{nd}}$ Stage Elec.  & 0.57    & 3.17    & -0.12  & -0.07   & 0.55  & -0.15  \\
2$^{\text{nd}}$ Stage Gas    & 10.51   & 11.05   & 10.19  & 0.94    & 0.97  & 0.89   \\ \bottomrule
\end{tabular}
\label{tab:cost_breakdown}
\end{table}

\section{Conclusion}
Gas-fired generation is a valuable source of flexibility in electric systems, and essential in managing variability and uncertainty from RES. However, this propagates RES uncertainty from the electric system into the gas system, and leads to an increased interdependence between the two infrastructures. In this paper, we formulate the gas-electric dispatch problem as a two-stage stochastic program with uncertain wind generation. The formulation accounts for the transient dynamic nature of natural gas system operations, which is important to accurately reflect the impact of uncertainty and short-term unbalanced gas operations. The drawback is that the resulting stochastic optimization problem is computationally challenging to solve. To address this challenge, we used a stochastic hybrid approximation algorithm. We proposed three different versions, based on different initial approximations. While all three formulations provide good results after tuning, we conclude that the approximation based on the certainty equivalent (SHACE) provides the best balance between ease of use (i.e. minimum tuning) and computational cost at each iteration. However, when appropriately tuned, the algorithm based on a simple convex approximation (SHACV) is faster and provides similarly good results. 
For the gas-electric scheduling problem, we were able to obtain solution times that is orders of magnitude faster than the benchmark approaches. Furthermore, we demonstrated that our algorithms achieve minimal load shed even in very challenging operational situations where scheduling without consideration of the natural gas system leads to significant load shed. 
In future work, we would like to pursue extended benchmarking and application of similar algorithms to other problems arising in coupled energy infrastructures.

%\appendices
%\section{} % no title needed
%\input{6_Appendix}
\ifCLASSOPTIONcaptionsoff
  \newpage
\fi

\bibliographystyle{IEEEtran}
\bibliography{ref.bib}

\begin{thebibliography}{10}
\providecommand{\url}[1]{#1}
\csname url@samestyle\endcsname
\providecommand{\newblock}{\relax}
\providecommand{\bibinfo}[2]{#2}
\providecommand{\BIBentrySTDinterwordspacing}{\spaceskip=0pt\relax}
\providecommand{\BIBentryALTinterwordstretchfactor}{4}
\providecommand{\BIBentryALTinterwordspacing}{\spaceskip=\fontdimen2\font plus
\BIBentryALTinterwordstretchfactor\fontdimen3\font minus
  \fontdimen4\font\relax}
\providecommand{\BIBforeignlanguage}[2]{{%
\expandafter\ifx\csname l@#1\endcsname\relax
\typeout{** WARNING: IEEEtran.bst: No hyphenation pattern has been}%
\typeout{** loaded for the language `#1'. Using the pattern for}%
\typeout{** the default language instead.}%
\else
\language=\csname l@#1\endcsname
\fi
#2}}
\providecommand{\BIBdecl}{\relax}
\BIBdecl

\bibitem{IEA}
{U.S. EIA}, ``Short-term energy outlook,'' 2020.

\bibitem{FERC_ER20}
``{FERC Proposal ER20-273-000},'' FERC, Tech. Rep., 2019.

\bibitem{omalley2020natural}
C.~O’Malley, S.~Delikaraoglou, L.~Roald, and G.~Hug, ``Natural gas system
  dispatch accounting for electricity side flexibility,'' \emph{Electric Power
  Systems Research}, vol. 178, p. 106038, 2020.

\bibitem{roald2020uncertainty}
L.~A. Roald, K.~Sundar, A.~Zlotnik, S.~Misra, and G.~Andersson, ``An
  uncertainty management framework for integrated gas-electric energy
  systems,'' \emph{arXiv preprint arXiv:2006.14561}, 2020.

\bibitem{omalley2018security}
C.~O'Malley, L.~Roald, D.~Kourounis, O.~Schenk, and G.~Hug, ``Security
  assessment in gas-electric networks,'' in \emph{2018 Power Systems
  Computation Conference (PSCC)}.\hskip 1em plus 0.5em minus 0.4em\relax IEEE,
  2018, pp. 1--7.

\bibitem{clegg2015integrated}
S.~Clegg and P.~Mancarella, ``Integrated electrical and gas network flexibility
  assessment in low-carbon multi-energy systems,'' \emph{IEEE Trans. on
  Sustainable Energy}, vol.~7, no.~2, pp. 718--731, 2015.

\bibitem{osiadacz1984simulation}
A.~Osiadacz, ``Simulation of transient gas flows in networks,'' \emph{Int. J.
  for numerical methods in fluids}, vol.~4, no.~1, pp. 13--24, 1984.

\bibitem{zlotnik2016coordinated}
A.~Zlotnik, L.~Roald, S.~Backhaus, M.~Chertkov, and G.~Andersson, ``Coordinated
  scheduling for interdependent electric power and natural gas
  infrastructures,'' \emph{IEEE Trans. on Power Syst.}, vol.~32, no.~1, pp.
  600--610, 2016.

\bibitem{mak2016efficient}
T.~W. Mak, P.~Van~Hentenryck, A.~Zlotnik, H.~Hijazi, and R.~Bent, ``Efficient
  dynamic compressor optimization in natural gas transmission systems,'' in
  \emph{ACC}.\hskip 1em plus 0.5em minus 0.4em\relax IEEE, 2016, pp.
  7484--7491.

\bibitem{correa2014gas}
C.~M. Correa-Posada and P.~S{\'a}nchez-Mart{\'\i}n, ``Gas network optimization:
  A comparison of piecewise linear models,'' \emph{Optimization Online}, 2014.

\bibitem{lohr2020optimal}
L.~L{\"o}hr, R.~Houben, and A.~Moser, ``Optimal power and gas flow for
  large-scale transmission systems,'' \emph{Elect. Power Syst. Research}, vol.
  189, p. 106724, 2020.

\bibitem{chaudry2008multi}
M.~Chaudry, N.~Jenkins, and G.~Strbac, ``Multi-time period combined gas and
  electricity network optimisation,'' \emph{Elect. power Syst. Research},
  vol.~78, no.~7, pp. 1265--1279, 2008.

\bibitem{tomasgard2007optimization}
A.~Tomasgard, F.~R{\o}mo, M.~Fodstad, and K.~Midthun, ``Optimization models for
  the natural gas value chain,'' in \emph{Geometric modelling, numerical
  simulation, and optimization}.\hskip 1em plus 0.5em minus 0.4em\relax
  Springer, 2007, pp. 521--558.

\bibitem{schwele2019coordination}
A.~Schwele, C.~Ordoudis, J.~Kazempour, and P.~Pinson, ``Coordination of power
  and natural gas systems: Convexification approaches for linepack modeling,''
  in \emph{2019 IEEE Milan PowerTech}.\hskip 1em plus 0.5em minus 0.4em\relax
  IEEE, 2019, pp. 1--6.

\bibitem{wang2017convex}
C.~Wang, W.~Wei, J.~Wang, L.~Bai, Y.~Liang, and T.~Bi, ``Convex optimization
  based distributed optimal gas-power flow calculation,'' \emph{IEEE Trans. on
  Sustainable Energy}, vol.~9, no.~3, pp. 1145--1156, 2017.

\bibitem{manshadi2018coordinated}
S.~D. Manshadi and M.~E. Khodayar, ``Coordinated operation of electricity and
  natural gas systems: a convex relaxation approach,'' \emph{IEEE Trans. on
  Smart Grid}, vol.~10, no.~3, pp. 3342--3354, 2018.

\bibitem{qadrdan2013operating}
M.~Qadrdan, J.~Wu, N.~Jenkins, and J.~Ekanayake, ``Operating strategies for a
  gb integrated gas and electricity network considering the uncertainty in wind
  power forecasts,'' \emph{IEEE Trans. on Sustainable Energy}, vol.~5, no.~1,
  pp. 128--138, 2013.

\bibitem{bertsimas2018robust}
D.~Bertsimas, V.~Gupta, and N.~Kallus, ``Robust sample average approximation,''
  \emph{Math. Program.}, vol. 171, no. 1-2, pp. 217--282, 2018.

\bibitem{liu2020dynamic}
K.~Liu, L.~T. Biegler, B.~Zhang, and Q.~Chen, ``Dynamic optimization of natural
  gas pipeline networks with demand and composition uncertainty,''
  \emph{Chemical Engineering Science}, vol. 215, p. 115449, 2020.

\bibitem{chen2019stochastic}
Z.~Chen, G.~Zhu, Y.~Zhang, T.~Ji, Z.~Liu, X.~Lin, and Z.~Cai, ``Stochastic
  dynamic economic dispatch of wind-integrated electricity and natural gas
  systems considering security risk constraints,'' \emph{CSEE J. of Power and
  Energy Syst.}, vol.~5, no.~3, pp. 324--334, 2019.

\bibitem{ratha2020affine}
A.~Ratha, A.~Schwele, J.~Kazempour, P.~Pinson, S.~S. Torbaghan, and A.~Virag,
  ``Affine policies for flexibility provision by natural gas networks to power
  systems,'' in \emph{Power Sys. Comp. Conf. (PSCC)}, 2020.

\bibitem{wang2019convex}
C.~Wang, W.~Wei, J.~Wang, and T.~Bi, ``Convex optimization based adjustable
  robust dispatch for integrated electric-gas systems considering gas delivery
  priority,'' \emph{Applied Energy}, vol. 239, pp. 70--82, 2019.

\bibitem{zhang2020distributionally}
Y.~Zhang, F.~Zheng, S.~Shu, J.~Le, and S.~Zhu, ``Distributionally robust
  optimization scheduling of electricity and natural gas integrated energy
  system considering confidence bands for probability density functions,''
  \emph{Int. J. of Elect. Power \& Energy Syst.}, vol. 123, p. 106321, 2020.

\bibitem{zhang2019two}
Y.~Zhang, J.~Le, F.~Zheng, Y.~Zhang, and K.~Liu, ``Two-stage distributionally
  robust coordinated scheduling for gas-electricity integrated energy system
  considering wind power uncertainty and reserve capacity configuration,''
  \emph{Renewable energy}, vol. 135, pp. 122--135, 2019.

\bibitem{zavala2014stochastic}
V.~M. Zavala, ``Stochastic optimal control model for natural gas networks,''
  \emph{Comput. Chem. Eng}, vol.~64, pp. 103--113, 2014.

\bibitem{zlotnik2016monotonicity}
A.~Zlotnik, S.~Misra, M.~Vuffray, and M.~Chertkov, ``Monotonicity of actuated
  flows on dissipative transport networks,'' in \emph{2016 European Control
  Conf. (ECC)}.\hskip 1em plus 0.5em minus 0.4em\relax IEEE, 2016, pp.
  831--836.

\bibitem{kekatos2014stochastic}
V.~Kekatos, G.~Wang, A.~J. Conejo, and G.~B. Giannakis, ``Stochastic reactive
  power management in microgrids with renewables,'' \emph{IEEE Trans. on Power
  Syst.}, vol.~30, no.~6, pp. 3386--3395, 2014.

\bibitem{kannan2020stochastic}
R.~Kannan, J.~R. Luedtke, and L.~A. Roald, ``Stochastic dc optimal power flow
  with reserve saturation,'' \emph{Elect. Power Syst. Research}, vol. 189, p.
  106566, 2020.

\bibitem{de2016adaptive}
T.~T. De~Rubira and G.~Hug, ``Adaptive certainty-equivalent approach for
  optimal generator dispatch under uncertainty,'' in \emph{2016 European
  Control Conf. (ECC)}.\hskip 1em plus 0.5em minus 0.4em\relax IEEE, 2016, pp.
  1215--1222.

\bibitem{cheung2000shape}
R.~K.-M. Cheung and W.~B. Powell, ``Shape--a stochastic hybrid approximation
  procedure for two-stage stochastic programs,'' \emph{Operations Research},
  vol.~48, no.~1, pp. 73--79, 2000.

\bibitem{nemirovski2009robust}
A.~Nemirovski, A.~Juditsky, G.~Lan, and A.~Shapiro, ``Robust stochastic
  approximation approach to stochastic programming,'' \emph{SIAM Journal on
  optimization}, vol.~19, no.~4, pp. 1574--1609, 2009.

\bibitem{yousefian2012stochastic}
F.~Yousefian, A.~Nedi{\'c}, and U.~V. Shanbhag, ``On stochastic gradient and
  subgradient methods with adaptive steplength sequences,'' \emph{Automatica},
  vol.~48, no.~1, pp. 56--67, 2012.

\bibitem{tan2016barzilai}
C.~Tan, S.~Ma, Y.-H. Dai, and Y.~Qian, ``Barzilai-borwein step size for
  stochastic gradient descent,'' in \emph{Advances in Neural Information
  Processing Syst.}, 2016, pp. 685--693.

\bibitem{taheri2020strategic}
S.~Taheri, V.~Kekatos, and H.~Veeramachaneni, ``Strategic investment in energy
  markets: A multiparametric programming approach,'' \emph{arXiv preprint
  arXiv:2004.06483}, 2020.

\bibitem{sahinidis1991convergence}
N.~Sahinidis and I.~E. Grossmann, ``Convergence properties of generalized
  benders decomposition,'' \emph{Comput. Chem. Eng}, vol.~15, no.~7, pp.
  481--491, 1991.

\bibitem{vandenbussche2019data}
B.~Vandenbussche, S.~Delikaraoglou, I.~Blanco, and G.~Hug, ``Data-driven
  adaptive benders decomposition for the stochastic unit commitment problem,''
  \emph{arXiv preprint arXiv:1912.01039}, 2019.

\bibitem{mazzibenders}
N.~Mazzi, A.~Grothey, K.~McKinnon, and N.~Sugishita, ``Benders decomposition
  with adaptive oracles for large scale optimization.''

\bibitem{bezanson2017julia}
\BIBentryALTinterwordspacing
J.~Bezanson, A.~Edelman, S.~Karpinski, and V.~B. Shah, ``Julia: A fresh
  approach to numerical computing,'' \emph{SIAM review}, vol.~59, no.~1, pp.
  65--98, 2017. [Online]. Available: \url{https://doi.org/10.1137/141000671}
\BIBentrySTDinterwordspacing

\bibitem{DunningHuchetteLubin2017}
I.~Dunning, J.~Huchette, and M.~Lubin, ``Jump: A modeling language for
  mathematical optimization,'' \emph{SIAM Review}, vol.~59, no.~2, pp.
  295--320, 2017.

\bibitem{wachter2006implementation}
A.~W{\"a}chter and L.~T. Biegler, ``On the implementation of an interior-point
  filter line-search algorithm for large-scale nonlinear programming,''
  \emph{Math. Prog.}, vol. 106, no.~1, pp. 25--57, 2006.

\bibitem{gurobi}
\BIBentryALTinterwordspacing
L.~Gurobi~Optimization, ``Gurobi optimizer reference manual,'' 2021. [Online].
  Available: \url{http://www.gurobi.com}
\BIBentrySTDinterwordspacing

\bibitem{data24}
``https://github.com/lanl-ansi/grail/blob/master/data/.''

\bibitem{bukhsh2015integrated}
W.~A. Bukhsh, C.~Zhang, and P.~Pinson, ``An integrated multiperiod opf model
  with demand response and renewable generation uncertainty,'' \emph{IEEE
  Trans. on Smart Grid}, vol.~7, no.~3, pp. 1495--1503, 2015.

\bibitem{winddata}
``https://sites.google.com/site/datasmopf/wind-scenarios.''

\end{thebibliography}

\end{document}